# Machinery Failure Approach and Spectral Analysis to study the Reaction Time Dynamics over Consecutive Visual Stimuli

M. E. Iglesias-Martínez[1,a] , M. Hernaiz-Guijarro[2,b], J. C. Castro-Palacio[2,c] , P. Fernández-de-Córdoba[2,d], J. M. Isidro[2,e], and E. Navarro-Pardo[3,f, *]

[1,] Departamento de Telecomunicaciones. Universidad de Pinar del Río, E-20100, Pinar del Río, Cuba.
a) migueliglesias2010@gmail.com, miigmar@doctor.upv.es

[2] Instituto Universitario de Matemática Pura y Aplicada, Grupo de Modelización Interdisciplinar, InterTech, Universitat Politècnica de València, E-46022, Valencia, Spain
  b) moihergu@doctor.upv.es c) juancas@upvnet.upv.es d) pfernandez@mat.upv.es e) joissan@mat.upv.es

[3] Departamento de Psicología Evolutiva y de la Educación. Grupo de Modelización Interdisciplinar, InterTech, Universitat de València, Valencia, E-46010, Spain
  f) esperanza.navarro@uv.es

* Correspondence: Esperanza Navarro Pardo, email: esperanza.navarro@uv.es

**Abstract**

The reaction times of individuals over consecutive visual stimuli have been studied using spectral analysis and a failure machinery approach. The used tools include the fast Fourier transform and a spectral entropy analysis. The results indicate that the reaction times produced by the independently responding individuals to visual stimuli appear to be correlated. The spectral analysis and the entropy of the spectrum yield that there are features of similarity in the response times of each participant and among them. Furthermore, the analysis of the mistakes made by the participants during the reaction time experiments concluded that they follow a behavior which is consistent with the MTBF (Mean Time Between Failures) model, widely used in industry for the predictive diagnosis of electrical machines and equipment.

1. Introduction

The literature relates an important number of works on human reaction times to visual stimuli [1, 2, 3, 4, 5, 6, 7, 8]. Fast responding is a very common scenario in daily life and in a broad variety of situations in industry [9], behavioral economics, finances [10], sports [11], and health [12], just to mention just a few examples.

The response time (RT) data have been proven to be correlated to cognitive disorders [8, 13, 14]. The most commonly diagnosed cognitive disorder in childhood affecting the RT is the Attention Deficit and Hyperactivity Disorder (ADHD). For instance, ADHD and autism spectrum disorder in children aged 7-10 years have been studied in reference [15] in order to gain insights into the attentional fluctuations, related to increased response time variability.

The fast Fourier transform (FFT) has been used to study the intra-individual variability (IIV) in children with ADHD. For instance, an FFT analysis of response time data in reference [16] revealed that there is a characteristic periodicity (frequency of 0.05 Hz) in children with ADHD. Similar studies yielded a periodicity at low frequency (0.03–0.07 Hz) observed in several tasks, namely, Sustained Attention to Response Task (SART) [17], the Ericksen Flanker Task [18], and the Go/NoGo Task [19].

On the other hand, error rates during response time experiments have been intensively studied in the literature as they retain important information about the neurological disorder under study. For children with ADHD, a relationship between error rates and the ex-Gaussian parameters of the response times has been found [20]. Similarly, increased error rates and long response times have been also found in developing readers and adults during yes/no and go/no-go tasks [21].

In this work, we will apply spectral analysis techniques such as the fast Fourier transform and entropy of the spectrum in order to get insights into the correlations and patterns in the response times of a group of individuals over consecutive visual stimuli. We will also perform a failure analysis of the mistakes made by the individuals while responding to the visual stimuli.

This article is organized as follows. In Section 2, the description of the sample (subsection 2.1) along with the response time experiments (subsection 2.2) are included. In Section 3, the methods used for the data analysis are described. The Results and Discussion are presented in Section 4, and the Conclusions in Section 5.

## 2. Description of the sample and visual stimuli experiments

### 2.1. Participants

A sample of 190 school-aged children from Valencia (Spain) took part in the computer-based experiments. The ages ranged from 8 to 10 years. Prior to the experiments, we made sure that all participants were healthy and had no brain injury, seizures, or any other neurological issue. The experiments were performed under the authorizations of the school management and the Regional Authority for Education. We have followed a protocol approved by the Government of Valencia. We also obtained the consent of the children's parents or legal guardians as stated by the Declaration of Helsinki [22].

## 2.2. Reaction time experiments

For the reaction time experiments, a child version of the Attentional Network Test (ANT Child) was chosen [23]. Every computerized experiment lasted for about 6-7 minutes in which 166 stimuli were presented randomly to the participants. The reaction times were recorded by DMDX [24], a Windows program widely used in the literature [8,25-27]. During each trial, 5 black fish in a white background and aligned horizontally were presented. The participants were asked to identify the right direction of the central fish by pressing the "M" key (the fish looking to the right) or "Z" otherwise. Every visual stimulus was presented for 2500 s. In general, Attention Network Tasks like the one in this work, pursue to assess three attentional networks: alerting, orienting, and executive control [28], which are closely related among themselves [29,30].

## 3. Methodology

For finite duration discrete-time signals $\{x(n)\}_{n=0}^{N-1}$ (which represents the signal to be processed) with *N* points of length, the classic method for estimation of the spectrum using the fast Fourier transform is defined as:

$$X(f) = \left| \sum_{n=0}^{N-1} x(n) e^{-j2\pi f \frac{n}{N}} \right| \quad (1)$$

where $f$ is the instantaneous frequency of the signal to be processed, $j$ is the imaginary unit, and $X(f)$ is the amplitude spectrum of the vector response times. The spectrum-based analysis is used as a supplement to the calculation of spectral entropy [31].

The entropy of a discrete-valued random variable attains a maximum value for a uniformly distributed variable. Considering a discrete random variable $Z$ with $M$ states $z_1, z_2, \ldots z_M$ and state probabilities $p_1, p_2, \ldots p_M$, that is $P(Z = z_i) = p_i$, the entropy of $Z$ is defined as:

$$H(Z) = -\sum_{i=1}^{M} p_i \log_2(p_i) \quad (2)$$

The spectral entropy of a signal is a measure of its spectral power distribution. The concept is based on Shannon's entropy. The spectral entropy treats the normalized energy distribution of the signal in the frequency domain as a probability distribution and calculates its Shannon entropy. This, the spectral entropy, $\mathrm{E}(F)$, is defined as [31]:

$$\mathrm{E}(F) = -\sum_{f=0}^{F} P(f) \cdot \log_2 P(f) \quad (3)$$

where $P(f)$ is the spectrum of the signal at the instantaneous frequency $f$.

## 4. Results and discussion

As a first analysis and to have a measure of the behavior of the response times throughout the entire data, the calculation of the mean value of the RTs to each stimulus over the sample of 190 children is performed. In other words, we construct a response time vector 166 (the number of items in the experiments) values of length, in which each component corresponds to the average response time of each visual stimulus over the sample of 190 participants. The results are shown in **Figure 1** with open circles. This time, the mistakes made by the participants while responding to the visual stimuli are not considered. This variable will be analyzed later.

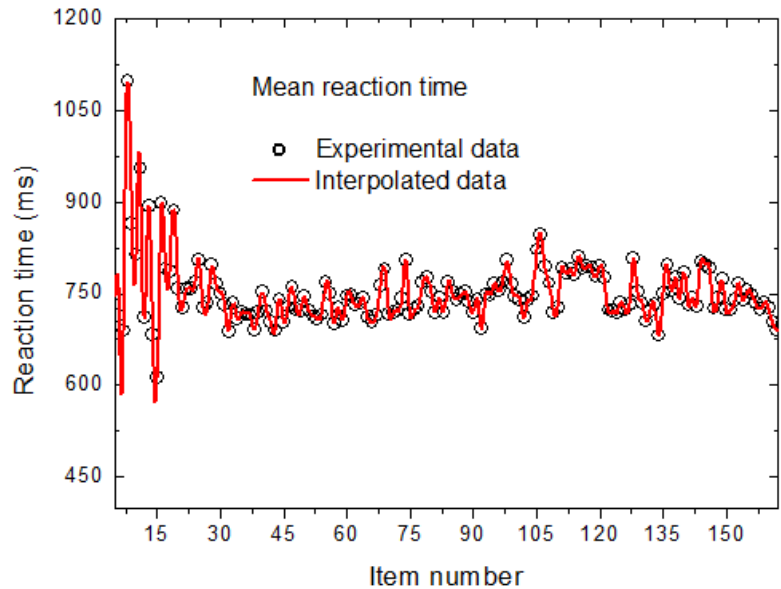

**Figure 1.** Mean reaction time over the sample of 190 participants *versus* the item ordering number.

For the purpose of completing the data of response time over consecutive stimuli for each participant, interpolations have been applied. Specifically, we have chosen an interpolation method in the framework of the reproducing kernel Hilbert space (RKHS) formalism which has been successfully applied in references [32-37] to represent the potential energy surface (PES) of small molecular systems. We decided to use this interpolation method as it shows several advantages over other methods, for instance, they are generic and parameter-free and for the interpolation at each point, the whole dataset is considered. By using the whole dataset, this method can retain any pattern manifested in the response time data over the consecutive stimuli.

This redensification of the data (upsampling) has been performed in order to increase the spectral resolution and to make the frequency spectrum look better visually. The original values are kept as well as their frequencies so that this process does not affect the original data. The results of the averaged RTs over the sample of 190 participants using this interpolation procedure are also presented in **Figure 1** with a solid red line.

The fast Fourier transform (FFT) technique has been applied to the response times of each participant in order to obtain the spectrum. **Figure 2** shows the average spectrum over the sample of participants. The average spectrum shows some frequency peaks which indicate that there are frequencies that are repetitive throughout the sample of all participants, which can result in a common spectral pattern. The main frequency of the spectrum is at 0.1 item$^{-1}$ which means that there is a common pattern every 10 items. The corresponding period of the pattern would be approximately 7.5 s, considering a mean value over the participants of 750 ms within the first 10 items (**Figure 1**). Studies of intra-individual variability in children have found that children with ADHD show a characteristic pattern in the response at a frequency of 0.05 Hz (every 20 s) *versus* a frequency of 0.075 Hz (13.3 s) in children taken as control sample [16].

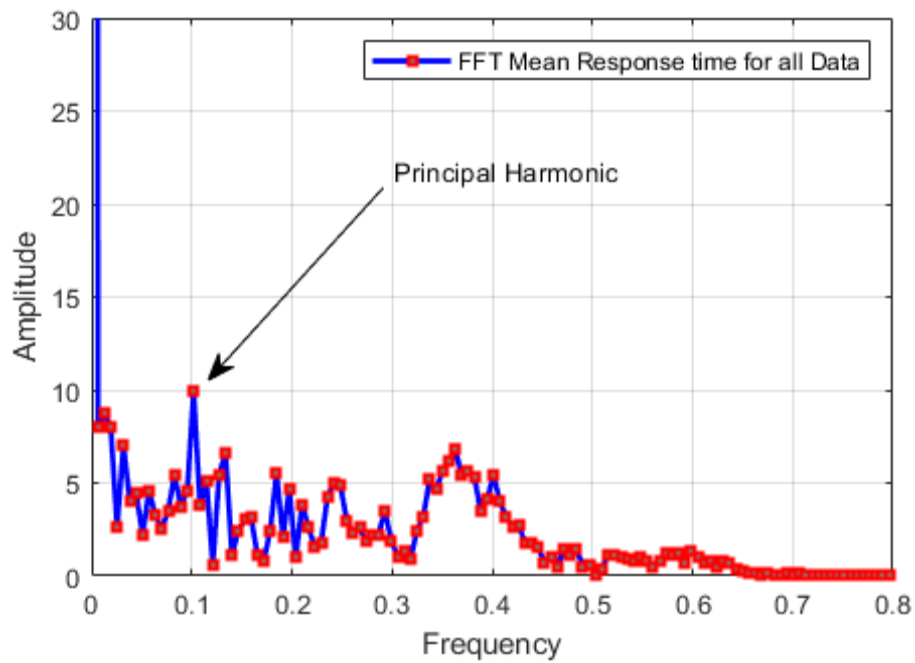

**Figure 2.** Mean FFT over a sample of 190 participants.

To complement the FFT results, the spectral entropy for the entire data sample is calculated. This calculation involves a spectral matrix of dimension 166 x 190 (166 reaction times for each of the 190 participants) which is evaluated to determine if there is a similar distribution in the frequency spectrum of each participant (equation (3)), what would imply a common reaction time against a common visual stimulus. To find out if there is a linear similarity relationship, the pairwise correlations across entire sample of students (190) are calculated. The correlation algorithm used was based on the Pearson correlation coefficient.

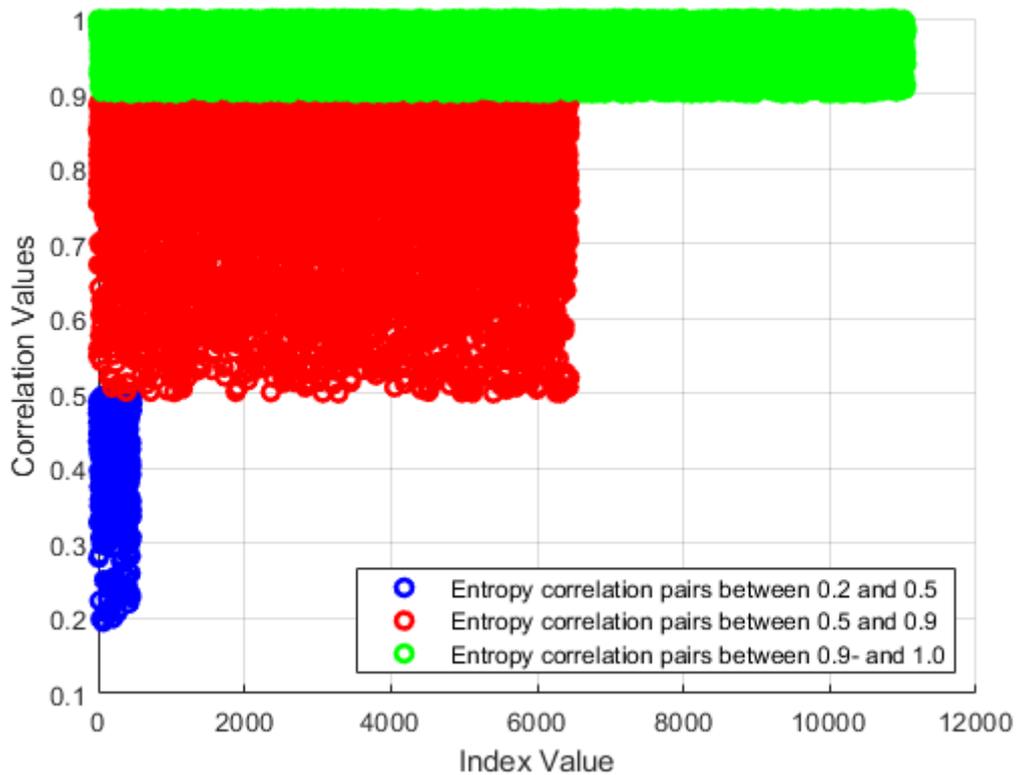

**Figure 3.** Spectral entropy correlation pairs grouped by intervals of association as a function of the pair index value within each group.

**Figure 3** shows the results of the correlation matrix calculated from the spectral entropy values over all participants. The graph includes a first group of 11026 correlation pair values greater than 0.9 (corresponding to a 61.41 % of the total number of pairs). This group involves a set of 149 participants with a very similar pattern of frequency distribution. There is a second group of 6441 correlations values (35.87 %) ranging between 0.5 and 0.9 and involving 114 participants presenting certain similar characteristics. Finally, a third group with 33 participants presented less distinctive similarities, with 528 correlations pairs values under 0.5 (2.94 %). Figure 3 provides an idea of the collective behavior in relation to independent characteristics of the group as it shows the intrinsic linear relationship between the response times of various individuals without sharing knowledge among them. This suggests some regularity and a certain common pattern for certain group characteristics in a population.

The analysis of the mistakes made by the participants while responding to the consecutive visual stimuli is made by calculating the mean spectral entropy over the sample of participants. **Figure 4** shows that the entropy value starts from a point of instability because the participants begin to adapt during the experimental process; then it follows a stationary behavior given that the participants are adapted to the visual stimuli showing a stable response time; then, at the end, they begin to visualize transient values and peaks in the entropy values, which is due to the fact that the participants have been running the process for a certain time and a dependent variable such as *"fatigue"*, can influence in the results.

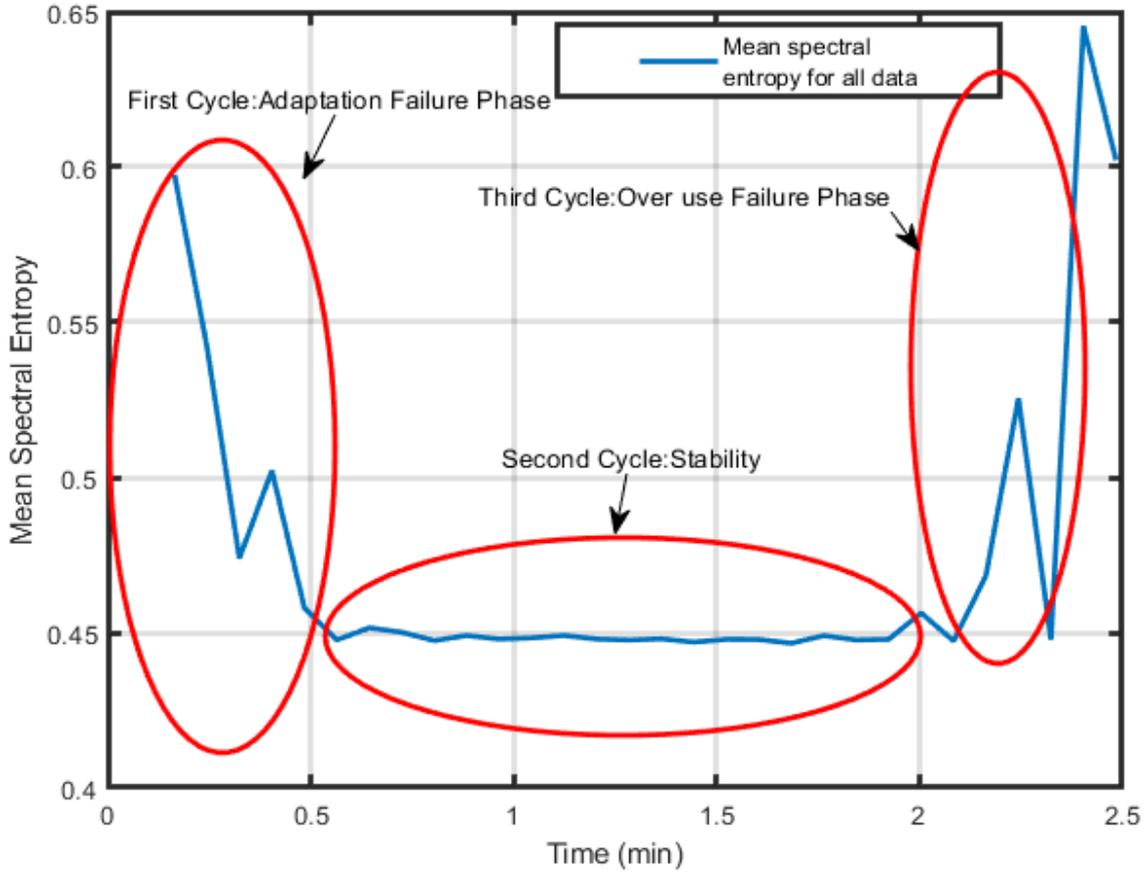

**Figure 4.** Mean spectral entropy as a function of time.

This result is in a good qualitative agreement with the failure modelling using Mean Time Between Failures (MTBF), widely used in industry for the predictive diagnosis of electrical machines and equipment. The mean time between failures is the arithmetic mean of the time between failures of a system. MTBF is typically part of a model that assumes that the failed system is repaired immediately as part of a renewal process [38]. The MTBF can be defined in terms of the expected value of the failure density function, namely $f(t)$:

$$MTBF : \int_0^\infty t\, f(t)\, dt, \qquad (4)$$

**Figure 5** illustrates a schematic representation of this model, where three stages are manifested. In the first stage, the machine is in the test period, when failures can occur randomly depending on the conditions and operating regimes. The second stage includes the stability period of the machine, when it has already passed the trial period and has adapted to the working conditions and regimes. Finally, the third stage refers to the period of overuse of the machine, time after the hours that the manufacturer guarantees for its proper functioning have passed, with which the failures can increase again.

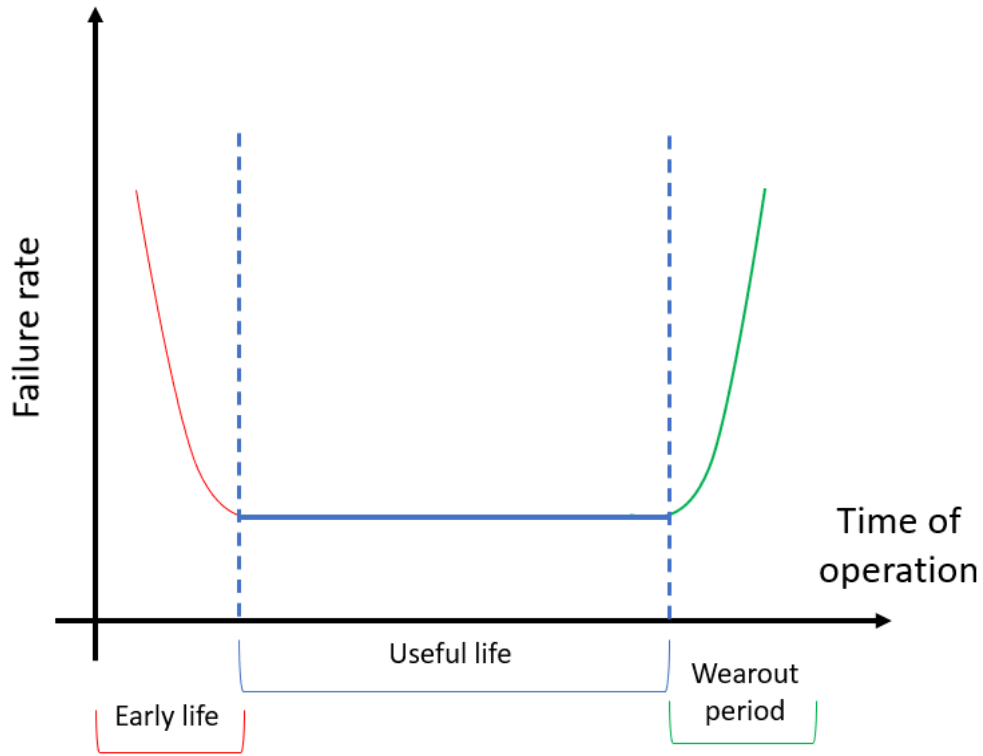

**Figure 5.** Schematic representation of the MTBF curve for machinery [39].

5.  **Conclusions**

Reaction times from visual stimuli experiments have been analyzed using a machinery failure approach and a spectral analysis. Results from the spectral analysis through the Fourier transform show that the participants have patterns in the responses along a series of visual stimuli, where the fundamental frequency for these experiments locates at 0.1 item$^{-1}$, that is, every 10 items (every around 7.5 s). Similarly, the correlation analysis based on the spectral entropy showed that there are correlations among the reaction time series of the participants who carry out the experiments independently and without exchanging information. This result can be rationalized by considering that the participants form a system or a collective whose responses are not independent from one another. On the other hand, the analysis of the mistakes made by the participants while responding to the visual stimuli shows that they follow a behavior, consistent with the MTBF model, used in the industry for the predictive diagnosis of electrical machines and equipment.